\documentclass[aps,prd,twocolumn,superscriptaddress,nofootinbib,10pt]{revtex4-2}

\usepackage[T1]{fontenc}
\usepackage{pslatex}
\usepackage{empheq}
\usepackage{amsmath}
\usepackage{amsfonts}
\usepackage{xcolor}
\usepackage{url}
\usepackage{bm}
\usepackage{graphicx}
\usepackage{amssymb}
\usepackage{verbatim}
\usepackage{soul}
\usepackage{booktabs}
\usepackage{array}
\usepackage[utf8]{inputenc}
\inputencoding{latin1}
\inputencoding{utf8}
\usepackage{appendix}

\newcommand{\araa}{ARA\&A}%
\newcommand{\apjl}{Astrophys.~J.}%
\newcommand{\ssr}{Space~Sci.~Rev.}%
\newcommand{\aap}{A\&A}%
\newcommand{\mnras}{MNRAS}%

\def\lsim{ \lower .75ex\hbox{$\sim$} \llap{\raise .27ex \hbox{$<$}} }
\def\gsim{ \lower .75ex \hbox{$\sim$} \llap{\raise .27ex \hbox{$>$}} }

\newcommand{\bi}{\begin{itemize}}
\newcommand{\ei}{\end{itemize}}

\usepackage[colorlinks=true,linkcolor=blue,citecolor=blue]{hyperref}

\begin{document}

\title{Polarization of synchrotron radiation from blazar jets}

\author{Filippo Bolis}
\email{filippo.bolis@inaf.it}
\affiliation{DiSAT, Università dell’Insubria, Via Valleggio 11, I-22100 Como, Italy}
\affiliation{INAF -- Osservatorio Astronomico di Brera, Via E. Bianchi 46, I-23807 Merate, Italy}

\author{Emanuele Sobacchi}
\affiliation{Gran Sasso Science Institute, Viale F.~Crispi 7, I-67100 L’Aquila, Italy}
\affiliation{INAF -- Osservatorio Astronomico di Brera, Via E. Bianchi 46, I-23807 Merate, Italy}

\author{Fabrizio Tavecchio}
\affiliation{INAF -- Osservatorio Astronomico di Brera, Via E. Bianchi 46, I-23807 Merate, Italy}

\begin{abstract}
Supermassive black holes in active galactic nuclei (AGNs) launch relativistic jets that shine through the entire electromagnetic spectrum. Blazars are a subclass of AGN where non-thermal radiation from the jet is strongly beamed, as the jet is directed nearly toward the observer. Multifrequency polarimetry is emerging as a powerful probe of blazar jets, especially with the advent of the Imaging X-ray Polarimetry Explorer (IXPE) space observatory. IXPE mostly targeted high synchrotron peaked (HSP) blazars, where both optical and X-ray emission can be attributed to synchrotron radiation from a population of non-thermal electrons. Observations of HSP blazars show that the polarization degree is strongly chromatic ($\Pi_{\rm X}/\Pi_{\rm O} \sim  2-7$), whereas the electric vector position angle (EVPA) is nearly independent of the observed frequency ($\Psi_{\rm X}\simeq\Psi_{\rm O}$). The strong chromaticity of the polarization degree was interpreted as an evidence that non-thermal electrons are accelerated by shocks. We present an alternative scenario that naturally explains IXPE observations. We study the polarization of synchrotron radiation from stationary axisymmetric jets viewed nearly on-axis. We show that the polarization degree increases significantly at high photon frequencies, as the distribution of the emitting electrons becomes softer, whereas the EVPA is nearly constant. The chromaticity of the polarization degree is much stronger in axisymmetric jets than in the case of a uniform magnetic field. Our results show that the topology of the electromagnetic fields is key to interpret multifrequency polarimetric observations of blazar jets. On the other hand, these observations may be less sensitive than previously thought to the specific particle acceleration process (e.g., shocks or magnetic reconnection).
\end{abstract}

\maketitle

\section{Introduction}
\label{sec:intro}

Relativistic jets launched by supermassive black holes in active galactic nuclei (AGNs) shine through the entire electromagnetic spectrum, from the radio band to very high-energy gamma-rays \cite{Blandford19}. Non-thermal radiation from AGN jets carries a wealth of information about the jet physics (e.g.,~non-thermal particle acceleration process, topology of the electromagnetic fields). Blazars, a subclass of AGNs where the jet is directed toward the observer, are ideal candidates to investigate the jet physics, as non-thermal radiation is strongly beamed due to the favorable jet orientation \cite{romero17, Boettcher19}.

Multifrequency polarimetry is emerging as a powerful probe of blazar jets, especially with the advent of the Imaging X-ray Polarimetry Explorer (IXPE) space observatory \cite{liodakis22, digesu22, digesu23, ehlert2023, Marshall2023, middei23b, peirson23, Errando2024, Kim2024}. Most multifrequency polarimetric observations target high synchrotron peaked (HSP) blazars, where both optical and X-ray emission can be attributed to synchrotron radiation from a population of non-thermal electrons. Observations of blazars in a quiescent state, where the polarization degree and the electric vector position angle (EVPA) do not show significant temporal variability, show that: (i)~The polarization degree, $\Pi$, is strongly chromatic. The X-ray polarization degree ($\Pi_{\rm X}\sim 10-20\%$) is significantly higher than the optical polarization degree ($\Pi_{\rm X}/\Pi_{\rm O}\sim 2-7$). On the other hand, (ii)~the EVPA, $\Psi$, is weakly chromatic ($\Psi_{\rm X} \simeq \Psi_{\rm O}$).\footnote{In the blazar Mrk 501, the EVPA was aligned with the projection of the jet axis on the plane of the sky \cite{liodakis22}. However the situation is far from clear, especially for the blazars 1ES 0229+200 \cite{ehlert2023} and Mrk 421 \cite{digesu22}. As emphasized by ref.~\cite{digesu22}, the jet could bend significantly between the inner region (where optical and X-ray emission are likely produced) and the outer region imaged in the radio band. Moreover, as shown by ref.~\cite{Kostrichkin}, jets can also change their direction in time.}

The strong dependence of the polarization degree on the observed frequency was interpreted as an evidence that non-thermal particles are accelerated by shocks \cite{liodakis22}. The argument in favor of this interpretation is the following \cite{tavecchio20, marscher21, tavecchio21}. Low energy optical emitting electrons have a longer cooling time than high energy X-ray emitting electrons, and therefore are advected and emit farther away from the shock front. If the jet magnetic field is turbulent, the polarization degree is expected to decrease as the emission region becomes more spatially extended, which would explain the observed chromaticity. 

The shock acceleration scenario challenges the leading theoretical paradigm for the launching of relativistic jets \cite{BlandfordZnajek1977, BlandfordPayne1982, Tchekhovskoy2011}. According to this paradigm, near the black hole jets are Poynting-dominated (i.e.,~most of their energy is carried by the Poynting flux). Part of the Poynting flux can be converted into bulk kinetic energy as the jet accelerates to relativistic bulk Lorentz factors, however the jet remains Poynting-dominated at the distance where the optical and X-ray emission are produced \cite{Lyubarsky2010}. Since shocks can hardly accelerate non-thermal particles in Poynting-dominated jets \cite{Sironi15b}, the shock acceleration scenario is problematic.

In our previous work \cite{Bolis+2024}, we presented an alternative scenario to interpret multifrequency polarimetric observations of HSP blazars. We studied the polarization of synchrotron radiation produced in axisymmetric stationary Poynting-dominated jets directed nearly toward the observer. The polarization degree and the EVPA are very sensitive  to the jet shape, which determines the global structure of the electromagnetic fields. Current observations can be reproduced if the jet is nearly parabolic, whereas a cylindrical jet shape is practically ruled out. The strong chromaticity of the polarization degree is solely due to the softening of the electron distribution at high energies. The dependence of the polarization degree on the slope of the electron distribution can be much stronger than for a uniform magnetic field.

In this paper, we follow a different, more general approach. We calculate approximate analytical expressions for the polarization degree and the EVPA of the synchrotron radiation from a generic axisymmetric stationary jet. The expressions that we derive are thus valid both for Poynting-dominated jets and matter-dominated jets (instead, results of our previous work are valid only for Poynting-dominated jets). An approximate analytical treatment is possible in the regime of small viewing angles, as appropriate for blazars. The paper is organized as follows. In Sec.~\ref{sec:stokes}, we present the general expressions for the Stokes parameters of the jet synchrotron radiation. In Sec.~\ref{sec:pol}, we calculate the polarization degree and the EVPA in the regime of small viewing angles. In Sec.~\ref{sec:application}, we apply our results to a specific model of Poynting-dominated jets. In Sec.~\ref{sec:disc}, we summarize our conclusions.

\section{Stokes parameters}
\label{sec:stokes}

We consider a stationary axisymmetric jet directed along the $z$ axis. The magnetic field can be presented as $\mathbf{B} = B_{R} \hat{\mathbf{R}} + B_{\phi} \hat{\bm{\phi}} +B_{z} \hat{\mathbf{z}}$. The bulk velocity of the fluid can be presented as $\mathbf{v} = v_{R} \hat{\mathbf{R}} + v_{\phi} \hat{\bm{\phi}} +v_{z} \hat{\mathbf{z}}$, and the corresponding Lorentz factor is $\Gamma=(1-v^2)^{-1/2}$. The electric field is given by $\mathbf{E} = - \mathbf{v}\times\mathbf{B}$, where the speed of light is set to $c=1$ (in stationary axisymmetric jets, the condition $E_\phi=0$ must be satisfied). In the proper frame of the fluid, the emitting electrons have a power-law energy distribution, namely
\begin{equation}
\frac{dN}{d\gamma} = K_{e} \gamma^{-p}\;, 
\end{equation}
where $K_{e}(R,z)$ is the proper electron number density, and $\gamma$ is the electron Lorentz factor. The distribution is assumed to be isotropic. 

The polarization of the synchrotron radiation from relativistic outflows was investigated by several authors \cite{BlandfordKoenigl79, Pariev2003, Lyutikov2003, Lyutikov05, DelZanna06, Bolis+2024}. 
The Stokes parameters per unit jet length and per unit transverse radius can be presented as
\begin{align}
\label{eq:stokes1gen}
Q & = \kappa_{p} \int_{0}^{2 \pi} {\rm d}\phi \; K_{e} \mathcal{D}^{\left( 3 + p\right)/2} \;  \left| \mathbf{B}' \times \hat{\mathbf{n}}' \right|^{\left( p+1 \right)/2} \; \cos 2 \chi \\
\label{eq:stokes2gen}
U & = \kappa_{p} \int_{0}^{2 \pi} {\rm d}\phi K_{e} \mathcal{D}^{\left( 3 + p\right)/2} \;  \left| \mathbf{B}' \times \hat{\mathbf{n}}' \right|^{\left( p+1 \right)/2}  \sin 2 \chi \; \\
\label{eq:stokes3gen}
I & = \frac{p + 7/3}{p+1} \; \kappa_{p} \int_{0}^{2 \pi} {\rm d}\phi \; K_{e} \mathcal{D}^{\left( 3 + p\right)/2} \;  \left| \mathbf{B}' \times \hat{\mathbf{n}}' \right|^{\left( p+1 \right)/2}\;,
\end{align}
where $\mathcal{D}$ is the Doppler factor, $|\mathbf{B}' \times \hat{\mathbf{n}}'|$ is the strength of the magnetic field component perpendicular to the line of sight, and $\chi$ is the angle between the polarization vector from a volume element and some reference direction in the plane of the sky that will be specified below. Primed quantities refer to the proper frame of the fluid. The constant $\kappa_p$, which does not affect the polarization degree and the EVPA, is defined by Eqs.~(12)-(13) of ref.~\cite{Bolis+2024}.

Following the framework developed by Del Zanna and collaborators \cite{DelZanna06}, we express the quantities in Eqs.~\eqref{eq:stokes1gen}-\eqref{eq:stokes3gen} as a function of the magnetic field and the bulk velocity of the fluid in the observer's frame. The unit vector directed toward the observer is $\hat{\mathbf{n}}= \sin{\theta_{\rm obs}}\cos{\phi}\; \hat{\mathbf{R}} - \sin{\theta_{\rm obs}} \sin{\phi}\; \hat{\bm{\phi}}+ \cos{\theta_{\rm obs}}\hat{\mathbf{z}}$, where the viewing angle $\theta_{\rm obs}$ is measured with respect to the direction of the jet axis, $\hat{\mathbf{z}}$. The polarization vector of the radiation from a volume element is \cite{Lyutikov2003, DelZanna06}
\begin{equation}
\label{eqn:polvec}
\hat{\mathbf{e}} = \frac{\hat{\mathbf{n}} \times \mathbf{q}}{\sqrt{  q^{2} - \left( \mathbf{q} \cdot \hat{\mathbf{n}} \right)^{2} }} \;,  \qquad \mathbf{q}=\mathbf{B} +\hat{\mathbf{n}}\times \left( \mathbf{v} \times \mathbf{B} \right) \;.
\end{equation}
We take as a reference direction the projection of the jet axis on the plane of the sky, which is $\hat{\mathbf{l}}=[(  \hat{\mathbf{z}} \cdot \hat{\mathbf{n}})\hat{\mathbf{n}}-\hat{\mathbf{z}}]/\sqrt{1-(\hat{\mathbf{z}} \cdot \hat{\mathbf{n}})^2}$ (note that $\hat{\mathbf{l}}$ is the unit vector orthogonal to $\hat{\mathbf{n}}$, and coplanar to $\hat{\mathbf{n}}$ and $\hat{\mathbf{z}}$). Then, the angle $\chi$ is determined by $\cos\chi=\hat{\mathbf{e}} \cdot \hat{\mathbf{l}}$ and $\sin\chi=\hat{\mathbf{e}} \cdot (\hat{\mathbf{l}} \times \hat{\mathbf{n}})$. It immediately follows that
\begin{equation}
\label{eqn:chi2}
\cos 2\chi = \frac{q_1^2-q_2^2}{q_1^2+q_2^2}\;, \qquad
\sin 2\chi = -\frac{2q_1q_2}{q_1^2+q_2^2} \;,
\end{equation}
where $q_1=\mathbf{q} \cdot (\hat{\mathbf{l}} \times \hat{\mathbf{n}})$ and $q_2=\mathbf{q} \cdot \hat{\mathbf{l}}$ are the two components of the projection of the vector $\mathbf{q}$ on the plane of the sky ($q_1$ and $q_2$ are respectively perpendicular and parallel to the projection of the jet axis on the plane of the sky). One finds
\begin{align}
\label{eq:q1}
q_1 & =  \left( E_{R} \cos \theta_{\rm obs} - B_{\phi} \right)\cos \phi - E_{z} \sin \theta_{\rm obs} - B_{R} \sin \phi \\
\label{eq:q2}
q_2 & = \left(E_{R} - B_{\phi} \cos \theta_{\rm obs} \right) \sin \phi - B_{z} \sin \theta_{\rm obs} + B_{R} \cos \theta_{\rm obs} \cos \phi \;.
\end{align}
The Doppler factor can be presented as
\begin{equation} 
\label{eq:doppler1}
\mathcal{D} = \frac{1}{\Gamma\left(1-v_3\right)}\;,
\end{equation}
where $v_3=\mathbf{v}\cdot\hat{\mathbf{n}}$ is the projection of the bulk velocity along the line of sight. One finds
\begin{equation}
 \label{eq:v3}
v_3 =  \left(v_{R}\cos \phi -v_{\phi}\sin \phi  \right)\sin \theta_{\rm obs} + v_{z}\cos \theta_{\rm obs} \;.
\end{equation}
As we demonstrate in Appendix \ref{sec:identities}, the strength of the magnetic field component perpendicular to the line of sight is
\begin{equation}
\label{eq:Bperp}
\left| \mathbf{B}' \times \hat{\mathbf{n}}' \right| = \mathcal{D} \sqrt{q_1^2+q_2^2} \;.  
\end{equation}
Substituting Eqs.~\eqref{eqn:chi2} and \eqref{eq:Bperp} into Eqs.~\eqref{eq:stokes1gen}-\eqref{eq:stokes3gen}, one can express the Stokes parameters as
\begin{align}
\label{stokes1}
    Q &= \kappa_{p}\int_{0} ^{2 \pi} d\phi \;K_{e}\mathcal{D}^{\left(p+2\right)} \left(q^{2}_{1} + q^{2}_{2} \right)^{\left(p-3\right)/4} \left( q^{2}_{1} - q^{2}_{2}\right) \\
    \label{stokes2}
     U &= \kappa_{p}\int_{0} ^{2 \pi} d\phi \;K_{e}\mathcal{D}^{\left(p+2\right)} \left(q^{2}_{1} + q^{2}_{2} \right)^{\left(p-3\right)/4} \left(-2q_{1}q_{2}\right) \\
     \label{stokes3}
    I &=\kappa_{p} \frac{p+7/3}{p+1}\int_{0} ^{2 \pi} d\phi \; K_{e}\mathcal{D}^{\left(p+2\right)} \left(q^{2}_{1} + q^{2}_{2} \right)^{\left(p+1\right)/4}\;.
\end{align}

\section{Polarization degree and EVPA}
\label{sec:pol}

The polarization degree of the synchrotron radiation from the jet (assumed to be unresolved) is given by
\begin{equation}
\label{PI}
\Pi = \frac{\sqrt{Q^{2} + U^{2}}}{I}\;. 
\end{equation}
The EVPA, $\Psi$, is given by
\begin{equation}
\label{eq:EVPA}
\cos 2 \Psi = \frac{Q}{\sqrt{Q^{2} + U^{2}}} \;, \qquad  \sin 2 \Psi = \frac{U}{\sqrt{Q^{2} + U^{2}}} \;.
\end{equation}
When the EVPA is parallel and perpendicular to the projection of the jet axis on the plane of the sky, one has respectively $\Psi=0$ and $\Psi=\pi/2$.

One could specify the jet electromagnetic fields, and then integrate Eqs.~\eqref{stokes1}-\eqref{stokes3} numerically to determine the Stokes parameters (this approach was followed by ref.~\cite{Bolis+2024}). Here we adopt a different approach, and calculate approximate expressions for $\Pi$ and $\Psi$ in the limit of small viewing angles, as appropriate for blazars. The main advantage of the latter approach is that the dependence of $\Pi$ and $\Psi$ on the physical parameters of the jet becomes transparent.


We expand all quantities in Eqs.~\eqref{stokes1}-\eqref{stokes3} as a power series of $\theta_{\rm obs}$, and then calculate the Stokes parameters at the first non-vanishing order. At the zeroth order (i.e.,~for $\theta_{\rm obs}=0$), one finds $Q=U=0$, whereas $I$ is given by
\begin{align}
\notag
I & = 2 \pi \kappa_{p} K_{e}\; \frac{p+7/3}{p+1}\\
\label{I}
&\quad \times\left[\frac{\sqrt{1-v^2}}{1-v_z}\right]^{p+2}\Big[ \left(B_\phi-E_R\right)^2+B_R^2 \Big]^{\left(p+1\right)/4}\;.
\end{align}
The first non-zero terms in the expansion of $Q$ and $U$ are proportional to $\theta_{\rm obs}^2$. One finds
\begin{align}
\label{Q}
Q & = \pi \kappa_{p} K_{e}\;
    \left[\frac{\sqrt{1-v^2}}{1-v_z}\right]^{p+2}
    \Big[ \left(B_\phi-E_R\right)^2+B_R^2 \Big]^{\left(p-3\right)/4}\widetilde{Q}\;\theta_{\rm obs}^2 \\
\label{U}
U & = \pi \kappa_{p} K_{e}\;
    \left[\frac{\sqrt{1-v^2}}{1-v_z}\right]^{p+2}
    \Big[ \left(B_\phi-E_R\right)^2+B_R^2 \Big]^{\left(p-3\right)/4}\widetilde{U}\;\theta_{\rm obs}^2
\end{align}
where we defined
\begin{align}
\notag
\widetilde{Q} &= \frac{p+5}{8}B^2\left(1-v^2+v_\parallel^2\right) +\frac{p^2-1}{16}\left[E_z^2 - B_z^2\right] \\
\notag
&\quad -\frac{p^2+3p+2}{4}B_z \Bigg[v_\parallel B-\frac{1-v^2}{1-v_z}B_z\Bigg] \\
\label{eq:uglyQ}
&\quad -\frac{p^2+5p+6}{4}\Bigg[v_\parallel B-\frac{1-v^2}{1-v_z}B_z\Bigg]^2 \\
\label{eq:uglyU} 
\widetilde{U} &= \frac{1-p^2}{8} E_z B_z -\frac{p^2+3p+2}{4}E_z\Bigg[v_\parallel B-\frac{1-v^2}{1-v_z}B_z\Bigg] \; .
\end{align}
We also defined $v_\parallel=\mathbf{v}\cdot\mathbf{B}/B$.

Substituting Eqs.~\eqref{I}-\eqref{U} into Eq.~\eqref{PI}, one finds
\begin{equation}
\label{poldegree}
\Pi = \frac{p+1}{p+7/3}\;\frac{\sqrt{\widetilde{Q}^{2} + \widetilde{U}^{2}}}{2 \left[ \left( B_{\phi} - E_{R} \right)^{2} + B^{2}_{R} \right] } \; \theta^{2}_{\rm obs}\;.
\end{equation}
Eq.~\eqref{poldegree} shows some important properties of the polarization degree of the synchrotron radiation from a generic jet: (i) $\Pi$ is proportional to $\Pi_{\rm max}=(p+1)/(p+7/3)$, which is the polarization degree obtained for a uniform magnetic field. (ii) $\Pi$ vanishes when the jet is viewed exactly on-axis (i.e.,~for $\theta_{\rm obs}=0$), as expected because the jet is axisymmetric. (iii) $\Pi$ has a strong dependence on the power-law index of the electron energy distribution, $p$, because $\widetilde{Q}^2 + \widetilde{U}^2$ is a polynomial of degree\footnote{The degree of the polynomial $\widetilde{Q}^2 + \widetilde{U}^2$ can be smaller than 4 for some specific jet models. For example, when $E_z=B_z=0$ and $v_\parallel=0$, Eqs.~\eqref{eq:uglyQ} and \eqref{eq:uglyU} give respectively $\widetilde{Q}=(p+5)(1-v^2)B^2/8$ and $\widetilde{U}=0$.} 4 in $p$. Then, the polarization degree increases much more rapidly than $\Pi_{\rm max}$ when the electron energy distribution becomes softer.

Substituting Eqs.~\eqref{Q}-\eqref{U} into Eq.~\eqref{eq:EVPA}, one finds
\begin{equation}\label{eq:psi}
\tan2\Psi = \frac{\widetilde{U}}{\widetilde{Q}}\;.
\end{equation}
Eq.~\eqref{eq:psi} shows some important properties of the EVPA: (i) $\Psi$ is independent of the viewing angle. (ii) $\Psi$ has a weak dependence on the power-law index of the electron energy distribution, $p$, because $\widetilde{U}/\widetilde{Q}$ is the ratio of polynomials of degree 2 in $p$. Then, the EVPA is nearly constant when the electron energy distribution becomes softer.

\section{Application to a specific jet model}
\label{sec:application}

In order to further simplify Eqs.~\eqref{poldegree}-\eqref{eq:psi}, one should consider a specific jet model. We assume that the jet is Poynting-dominated, consistent with the leading theoretical paradigm for the launching of AGN jets \cite{BlandfordZnajek1977, BlandfordPayne1982, Tchekhovskoy2011}. The structure of Poynting-dominated jets was thoroughly investigated both numerically \cite{Komissarov2007, Komissarov2009, Tchekhovskoy2008, Tchekhovskoy09} and analytically \cite{Beskin+1998, Beskin+2004, Vlahakis2004, Lyubarsky2009, Lyubarsky2010, Lyubarsky2011}. We use the model of Lyubarsky \cite{Lyubarsky2009}, which describes axisymmetric, stationary, relativistic, Poynting-dominated outflows collimated by an external medium. 

We assume that the confining pressure can be parametrized as $\mathcal{P}_{\rm ext} \propto z_0^{-\kappa}$, where $z_0$ is the distance from the black hole. In this case, the jet transverse radius is $R_{0}\propto z_{0}^{q}$. For $0<\kappa<2$ one has $q=\kappa/4$, whereas for a wind-like medium with $\kappa=2$ one has $1/2<q<1$. For $\kappa=2$, the value of $q$ depends on the ratio of the magnetic and external pressure near the light cylinder. The jet electromagnetic fields can be presented as
\begin{align}
\label{eq:Ecomp}
E_{R} & = \Omega R B_{\rm p} \cos \Theta\;, & E_{z} & = -\Omega R B_{\rm p} \sin \Theta \\
B_{R} & = B_{\rm p}\sin \Theta\;, & B_{z} & = B_{\rm p} \cos \Theta \;.
\end{align}
The poloidal magnetic field, $B_{\rm p}$, and the angular velocity of the field lines, $\Omega$, are assumed to be independent of $R$. The local jet opening angle, $\Theta$, is determined by the shape of the magnetic flux surfaces. As discussed by ref.~\cite{Bolis+2024}, one has
\begin{equation}
\label{eq:opAngle}
 \Theta =  q \; \frac{R}{R_{0}} \frac{3^{1 / 4q}}{\left(\Omega R_{0} \right)^{\left(1-q\right) / q}}\;.
\end{equation}
The toroidal magnetic field is determined by
\begin{equation}
\label{eq:BminusE}
 \frac{B^{2}_{\phi} - E^{2}}{B^{2}_{\rm p}} = \frac{q \left(1-q \right)}{3^{1-1/2q}}  \frac{\left(\Omega R\right)^{4}}{\left(\Omega R_{0}\right)^{2 / q}} \;.
\end{equation}
We assume that the bulk velocity of the fluid, $\mathbf{v}$, is equal to the drift velocity, $\mathbf{E}\times\mathbf{B}/B^2$, which implies $v_\parallel=0$. Then, the bulk Lorentz factor is $\Gamma=(1-E^2/B^2)^{-1/2}$.

We consider electrons that fill an annulus near the edge of the jet, which is located in $R=R_{0}$. This approximation is justified when the jet magnetization is constant, as the electron number density peaks near the jet boundary \cite{Lyubarsky2009, Bolis+2024}. When the jet is significantly accelerated, one has $\Omega R_0\gg 1$. Retaining only the highest power of $\Omega R_0$, the bulk Lorentz factor of the annulus, $\Gamma_0=\Gamma(R_0)$, can be approximated as
\begin{empheq}[left={\Gamma_0 \simeq\empheqlbrace}]{align}
   &\Omega R_0  & {\rm for\;\;} 0<q<1/2 \label{eq:gamma0cyl}\\
   & \sqrt{\frac{3^{1-1/2q}}{q\left(1-q\right)}}\left(\Omega R_0\right)^{-1 +1/q} & {\rm for\;\;} 1/2<q<1 \label{eq:gamma0parab}
  \end{empheq}
One can also expand the expressions for the electromagnetic fields, retaining only the highest power of $\Omega R_0$.
For example, one can approximate
\begin{align}
\label{approximation1}
\frac{E_z}{B_{\rm p}} & \simeq -3^{1/4q}q\left(\Omega R_0\right)^{2-1/q} \\
\label{approximation2}
\frac{B_R}{B_{\rm p}} & \simeq 3^{ 1/4q}q (\Omega R_0)^{1-1/q} \\
\label{approximation3}
\frac{B_{\phi} -  E_{R}}{B_{\rm p}} & \simeq 3^{ 1/2q}\; \frac{q\left(2q+1\right)}{6} (\Omega R_{0})^{3 - 2/q} \;.
\end{align}
Below we determine approximate expressions for the polarization degree and the EVPA, which depend on the jet shape through the parameter $q$. We consider nearly cylindrical jets ($0<q<1/2$) and nearly parabolic jets ($1/2<q<1$).

\subsection{Nearly cylindrical jets ($\mathbf{0<q<1/2}$)}

We consider nearly cylindrical jets where $0<q<1/2$. In the limit of large Lorentz factors ($\Omega R_0\gg 1$), one can retain only the highest power of $\Omega R_0$. Then, Eqs.~\eqref{eq:uglyQ}-\eqref{eq:uglyU} can be approximated as
\begin{align}
\label{qcyl}
\widetilde{Q} &\simeq -\frac{\left(p+1\right)\left(p+5\right)}{16} B_{\rm p}^2 \\
\label{ucyl}
\widetilde{U} &\simeq -\frac{\left(p+1\right)\left(p+5\right)}{8} 3^{1/4q}q \left( \Omega R_{0}\right)^{2 - 1/q} B_{\rm p}^2 \;.
\end{align}

Using Eq.~\eqref{eq:gamma0cyl}, one can express $\Omega R_0$ as a function of the bulk Lorentz factor, $\Gamma_0$. Since $(\Omega R_0)^{2-1/q}\ll 1$ for $q<1/2$, Eqs.~\eqref{approximation2}-\eqref{approximation3} imply $(B_\phi-E_R)^2\ll B_R^2$, and Eqs.~\eqref{qcyl}-\eqref{ucyl} imply $|\widetilde{Q}| \gg |\widetilde{U}|$. Then, Eq.~\eqref{poldegree} gives
\begin{equation}
\label{polcyl}
\Pi \simeq \frac{\left(p+1\right)^{2} \left( p+ 5 \right)}{p+7/3}\; \frac{3^{-1/2q}}{32 q^{2}} \Gamma_{0}^{-2 + 2/q} \;\theta^{2}_{\rm obs}\;.
\end{equation}
Eq.~\eqref{polcyl} shows that the polarization degree increases rapidly when the jet is viewed slightly off-axis. For example, one finds $\Pi\sim\Pi_{\rm max}$ for $\theta_{\rm obs}\sim \Gamma_0^{1-1/q}$. For $q<1/2$, the latter viewing angle is much smaller than the typical viewing angle of blazar jets, which is on the order of $1/\Gamma_0$. Eq.~\eqref{polcyl} is an accurate approximation of the polarization degree when $\Pi\ll\Pi_{\rm max}$, whereas it overestimates the polarization degree by a factor of a few when $\Pi\sim\Pi_{\rm max}$ (see Appendix \ref{sec:val}).

Since $|\widetilde{Q}| \gg |\widetilde{U}|$ and $\widetilde{Q}<0$, the EVPA is nearly perpendicular to the projection of the jet axis on the plane of the sky. A small misalignment with respect to $\Psi=\pi/2$ can be calculated using Eq.~\eqref{eq:psi}, which gives
\begin{equation}
\label{EVPAcyl}
\Psi \simeq \frac{\pi}{2} +  3^{1/4q}q\Gamma_{0}^{2- 1/q}\;.
\end{equation}

Eqs.~\eqref{polcyl}-\eqref{EVPAcyl} suggest that in blazars (where the typical viewing angle is $\theta_{\rm obs}\sim 1/\Gamma_0$) one has $\Pi\sim\Pi_{\rm max}$ and $\Psi\sim\pi/2$, consistent with our previous results obtained by integrating Eqs.~\eqref{stokes1}-\eqref{stokes3} numerically \cite{Bolis+2024}. Then, a nearly cylindrical jet shape is practically ruled out by current multifrequency polarimetric observations of HSP blazars, as these observations show that $\Pi<\Pi_{\rm max}$ and $\Psi\sim 0$ \cite{liodakis22}.

\subsection{Nearly parabolic jets ($\mathbf{1/2<q<1}$)}

We consider nearly parabolic jets where $1/2<q<1$. In the limit of large Lorentz factors ($\Omega R_0\gg 1$), Eqs.~\eqref{eq:uglyQ}-\eqref{eq:uglyU} can be approximated as
\begin{align}
\label{qpar}
\widetilde{Q} &\simeq  \frac{q\;3^{1/2q}}{48} \Big[3qp^{2} + 2p\left(1-q \right) -13q +10 \Big] \left( \Omega R_{0}\right)^{4 - 2/q} B_{\rm p}^2 \\
\label{upar}
\widetilde{U} &\simeq \frac{\left(p+1\right)}{8} \frac{q\;3^{1+1/4q}}{\left(2q+1 \right)} \Big[p\left(2q -1\right) + 2q -3 \Big]\left( \Omega R_{0}\right)^{2 - 1/q} B_{\rm p}^2 \;.
\end{align}

Using Eq.~\eqref{eq:gamma0parab}, one can express $\Omega R_0$ as a function of the bulk Lorentz factor, $\Gamma_0$. Since $(\Omega R_0)^{2-1/q}\gg 1$ for $q>1/2$, Eqs.~\eqref{approximation2}-\eqref{approximation3} imply $(B_\phi-E_R)^2\gg B_R^2$, and Eqs.~\eqref{qpar}-\eqref{upar} imply $|\widetilde{Q}| \gg |\widetilde{U}|$. Then, Eq.~\eqref{poldegree} gives
\begin{align}
\notag
\Pi & \simeq \frac{p+1}{p+7/3}\; \frac{\left(1-q\right)}{8\left(2q+1\right)^{2}} \\ 
\label{polparab}
&\quad \times \Big[ 3qp^{2} + 2p\left(1-q \right) -13q +10   \Big]\; \Gamma^{2}_{0} \; \theta^{2}_{\rm obs} \;.
\end{align}
Eq.~\eqref{polparab} shows that $\Pi$ is significantly smaller than $\Pi_{\rm max}$ for large jet opening angles (i.e.,~$|1-q|\ll 1$), because $\Pi$ is proportional to $1-q$. As we discussed earlier, $\Pi$ increases much more rapidly than $\Pi_{\rm max}$ when the electron energy distribution becomes softer, because $\Pi$ is proportional to a polynomial of degree 2 in $p$. For example, in a jet where $q=0.7$ and $\Gamma_0\theta_{\rm obs}=1$, one finds $\Pi\simeq 5\%$ for $p=2$, and $\Pi\simeq 20\%$ for $p=4$ (instead, one has $\Pi_{\rm max}=70\%$ for $p=2$, and $\Pi_{\rm max}=80\%$ for $p=4$). Eq.~\eqref{polparab} is an accurate approximation of the polarization degree for all the viewing angles relevant for blazars, namely $\theta_{\rm obs}\lesssim 1/\Gamma_0$ (see Appendix \ref{sec:val}).

Since $|\widetilde{Q}| \gg |\widetilde{U}|$ and\footnote{One can show that $3qp^{2} + 2p\left(1-q \right) -13q +10>0$ for $1/2<q<1$ and $p>1$. Since in blazars the power-law index of the electron energy distribution is softer than $p=1$, from Eq.~\eqref{qpar} one finds $\widetilde{Q}>0$.} $\widetilde{Q}>0$, the EVPA is nearly parallel to the projection of the jet axis on the plane of the sky. A small misalignment with respect to $\Psi=0$ can be calculated using Eq.~\eqref{eq:psi}, which gives
\begin{align}
\notag
\Psi & \simeq \frac{3^{2-1/4q}}{2q+1}\; \Bigg[\frac{q\left(1-q\right)}{3^{1-1/2q}}\Bigg]^{\left( 1-2q\right)/2\left( 1-q\right)} \\
\label{EVPAparab}
&\quad \times \frac{p^{2}\left(2q -1\right) + 4p\left(q -1\right) + 2q -3}{3qp^{2} + 2p\left(1-q \right) -13q +10} \; \Gamma_{0}^{\left( 1-2q\right)/\left( 1-q\right) }
\end{align}
For example, in a jet where $q=0.7$ and $\Gamma_0=10$, one finds $\Psi\simeq 0.1{\rm\; rad}$ for $p=2$, and $\Psi\simeq 0{\rm\; rad}$ for $p=4$.

Eqs.~\eqref{polparab}-\eqref{EVPAparab} show that nearly parabolic jets can explain multifrequency polarimetric observations of HSP blazars, where $\Pi<\Pi_{\rm max}$ and $\Psi\sim 0$ \cite{liodakis22}. The fact that the X-rays are significantly more polarized than the optical radiation can be attributed to the strong dependence of $\Pi$ on the power-law index of the electron energy distribution, $p$.

\section{Conclusions}
\label{sec:disc}

We studied the polarization of synchrotron radiation from blazar jets. We derived approximate analytical expressions for the polarization degree, $\Pi$, and the EVPA, $\Psi$, of a generic axisymmetric stationary jet. In the limit of small viewing angles, $\Pi$ and $\Psi$ are given respectively by Eqs.~\eqref{poldegree} and \eqref{eq:psi}. We found that:
\begin{itemize}
\item The polarization degree, $\Pi$, has a strong dependence on $p$ (where $p$ is the power-law index of the electron energy distribution), because $\Pi$ is proportional to $\Pi_{\rm max}=(p+1)/(p+7/3)$, times the square root of a polynomial of degree 4 in $p$. Then, $\Pi$ increases much more rapidly than $\Pi_{\rm max}$ when the electron energy distribution becomes softer (i.e.,~for large $p$).
\item  The EVPA, $\Psi$, has a weak dependence on $p$, because $\tan2\Psi$ is the ratio of polynomials of degree 2 in $p$. Then, $\Psi$ is nearly constant when the electron energy distribution becomes softer.
\end{itemize}
Our results are important to interpret multifrequency polarimetric observations of HSP blazars \cite{liodakis22, digesu22, digesu23, ehlert2023, Marshall2023, middei23b, peirson23, Errando2024, Kim2024}. Observations show that X-rays are significantly more polarized than optical radiation ($\Pi_{\rm X}/\Pi_{\rm O} \sim  2-7$), whereas the EVPA does not change significantly ($\Psi_{\rm X}\simeq\Psi_{\rm O}$). The strong chromaticity of the polarization degree, and the weak chromaticity of the EVPA, can be attributed to the fact that X-ray emitting electrons have a softer energy distribution than optical emitting electrons (X-ray and optical emitting electrons have respectively $p=4-6$ and $p=2$ \cite{fossati98, Abdo2011, ghisellini17}). We emphasize that results discussed so far are independent of the specific jet model and particle acceleration mechanism.\footnote{Fig.~5 of ref.~\cite{Lyutikov2003} and Fig.~3 of ref.~\cite{Bolis+2024} show that $\Pi$ can increase much more rapidly than $\Pi_{\rm max}$ for large $p$. Fig.~3 of ref.~\cite{Bolis+2024} shows that $\Psi$ is nearly independent of $p$. These results were obtained by considering specific jet models, whereas in this paper we considered a generic axisymmetric stationary jet.}

We considered the model of Poynting-dominated jets developed by Lyubarsky \cite{Lyubarsky2009}. In this model, the electromagnetic fields are determined by the jet shape (the jet transverse radius is $R_0\propto z_0^q$, where $z_0$ is the distance from the black hole, and $0<q<1$). We derived approximate analytical expressions for $\Pi$ and $\Psi$, in the limit of large bulk Lorentz factors (i.e.,~$\Gamma_0\gg 1$):
\begin{itemize}
\item For nearly cylindrical jets ($0<q<1/2$), $\Pi$ and $\Psi$ are given respectively by Eqs.~\eqref{polcyl} and \eqref{EVPAcyl}. The polarization degree becomes comparable with $\Pi_{\rm max}$ for viewing angles $\theta_{\rm obs} \ll 1/\Gamma_0$, and the EVPA is nearly perpendicular to the projection of the jet axis on the plane of the sky (Eq.~\ref{polcyl} provides an accurate approximation of the polarization degree when $\Pi\ll\Pi_{\rm max}$, whereas it overestimates the polarization degree by a factor of a few when $\Pi\sim\Pi_{\rm max}$).
\item For nearly parabolic jets ($1/2<q<1$), $\Pi$ and $\Psi$ are given respectively by Eqs.~\eqref{polparab} and \eqref{EVPAparab}. The polarization degree is significantly smaller than $\Pi_{\rm max}$ for viewing angles $\theta_{\rm obs}\lesssim 1/\Gamma_0$, and the EVPA is nearly parallel to the projection of the jet axis on the plane of the sky (Eqs.~\ref{polparab} and \ref{EVPAparab} provide an accurate approximation of the polarization degree and of the EVPA for viewing angles $\theta_{\rm obs}\lesssim 1/\Gamma_0$).
\end{itemize}
Multifrequency polarimetric observations of HSP blazars (where $\theta_{\rm obs}\sim 1/\Gamma_0$) show that $\Pi<\Pi_{\rm max}$, and $\Psi\sim 0$ \cite{liodakis22}. Then, as previously discussed by ref.~\cite{Bolis+2024}, Poynting-dominated jets should be nearly parabolic, whereas a cylindrical shape is practically ruled out.

\begin{acknowledgements}
We acknowledge financial support from the Marie Sk{\l}odowska-Curie Grant 101061217 (PI E.~Sobacchi) and from a INAF Theory Grant 2022 (PI F.~Tavecchio). This work was funded by the European Union-Next Generation EU, PRIN 2022 RFF M4C21.1 (2022C9TNNX).
\end{acknowledgements}

\appendix

\begin{figure*}[htb]
\centering
\includegraphics[width=0.48\linewidth]{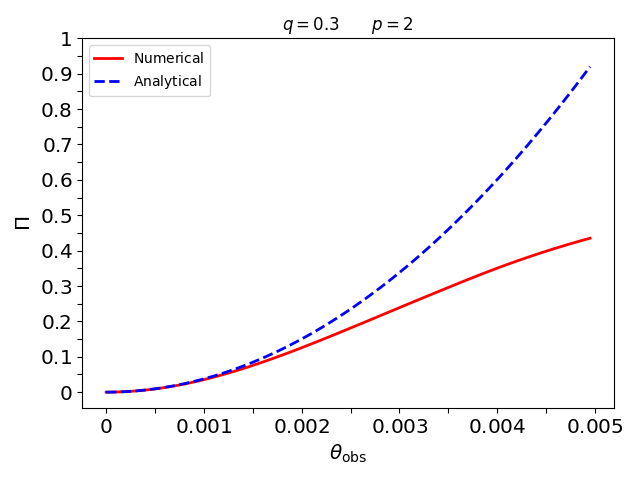}
\includegraphics[width=0.48\linewidth]{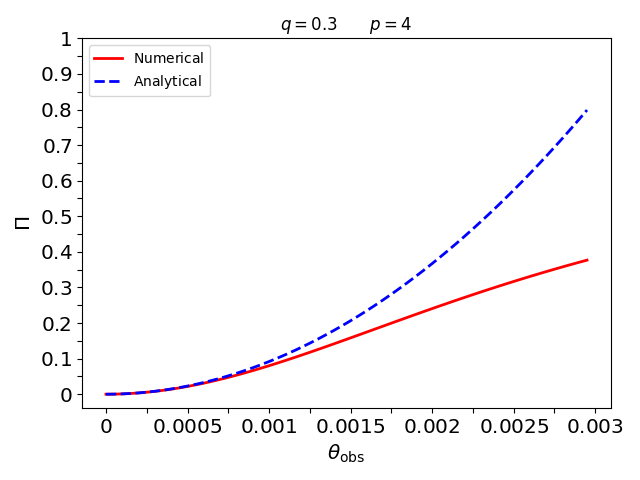}
\caption{Polarization degree, $\Pi$, as a function of the viewing angle, $\theta_{\rm obs}$ (expressed in radians), for a a nearly cylindrical jet where $q=0.3$. The bulk Lorentz factor is $\Gamma_0=10$. The power-law index of the electron energy distribution is $p=2$ (left panel) and $p=4$ (right panel). The solid line is obtained by integrating Eqs.~\eqref{stokes1}-\eqref{stokes3} numerically (we evaluated the electromagnetic fields in $R=R_0$, where the edge of the jet is located), whereas the dashed line corresponds to our analytical approximation (Eq.~\ref{polcyl}). The scale of the $x$ axis is different in the two panels.}
\label{fig:q03}
\end{figure*}
\begin{figure*}[htb]
\centering
\includegraphics[width=0.48\linewidth]{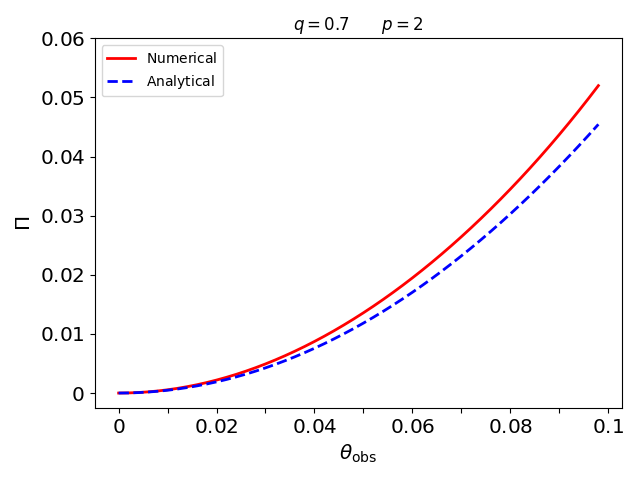}
\includegraphics[width=0.48\linewidth]{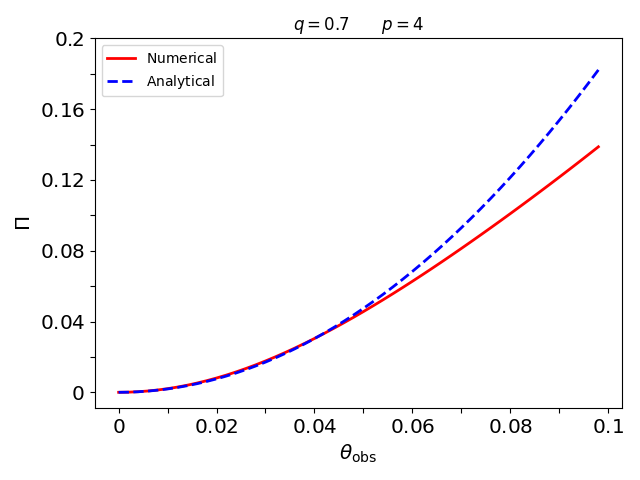}
\caption{Same as Fig.~\ref{fig:q03}, for a nearly parabolic jet where $q=0.7$. The dashed line corresponds to our analytical approximation (Eq.~\ref{polparab}). The $x$ axis extends to $\theta_{\rm obs}=1/\Gamma_0$. The scale of the $y$ axis is different in the two panels.}
\label{fig:q07}
\end{figure*}

\section{Perpendicular magnetic field}
\label{sec:identities}

Here we derive Eq.~\eqref{eq:Bperp}.  
The vector $\mathbf{q}=\mathbf{B} +\hat{\mathbf{n}}\times \left( \mathbf{v} \times \mathbf{B} \right)$ can be presented as
\begin{equation}
\label{eq:A1}
\mathbf{q} = \mathbf{B} +\left(\mathbf{B}\cdot\hat{\mathbf{n}}\right)\mathbf{v} -\left( \mathbf{v}\cdot \hat{\mathbf{n}} \right) \mathbf{B}\;.
\end{equation}
From Eq.~\eqref{eq:A1}, one finds
\begin{align}
\notag
q^2 - \left(\mathbf{q}\cdot\hat{\mathbf{n}}\right)^2 & =B^2 (1-\mathbf{v} \cdot \hat{\mathbf{n}} )^2 - (\mathbf{B} \cdot\hat{\mathbf{n}})^2 (1-v^2) \\
& \quad + 2 (\mathbf{B} \cdot\hat{\mathbf{n}}) ( \mathbf{v} \cdot \mathbf{B} )(1-\mathbf{v} \cdot \hat{\mathbf{n}} )\;.
\end{align}
The latter expression can be presented as
\begin{equation}
\label{eq:A2}
q^2 - \left(\mathbf{q}\cdot\hat{\mathbf{n}}\right)^2 = \frac{B^2}{\Gamma^2 \mathcal{D}^2}-\frac{(\mathbf{B}\cdot\hat{\mathbf{n}})^2}{\Gamma^2} + 2 \frac{(\mathbf{B} \cdot\hat{\mathbf{n}}) (\mathbf{B} \cdot \mathbf{v})}{ \Gamma\mathcal{D} }\;.
\end{equation}
The strength of the magnetic field component perpendicular to the line of sight is given by \cite{DelZanna06}
\begin{equation}
\label{eq:A3}
| \mathbf{B}' \times \hat{\mathbf{n}}'|  =\mathcal{D}\sqrt{\frac{B^2}{\Gamma^2 \mathcal{D}^2}-\frac{(\mathbf{B} \cdot\hat{\mathbf{n}})^2}{\Gamma^2}
+ 2 \frac{(\mathbf{B} \cdot\hat{\mathbf{n}}) (\mathbf{B} \cdot \mathbf{v})}{ \Gamma \mathcal{D} }} \;. 
\end{equation}
Comparing Eqs.~\eqref{eq:A2} and \eqref{eq:A3}, one finds
\begin{equation}
\left| \mathbf{B}' \times \hat{\mathbf{n}}' \right| = \mathcal{D} \sqrt{q^2 - \left(\mathbf{q}\cdot\hat{\mathbf{n}}\right)^2} \;,  
\end{equation}
which is equivalent to Eq.~\eqref{eq:Bperp}.

\section{Validity of the analytical approximation}
\label{sec:val}

In Figs.~\ref{fig:q03} and \ref{fig:q07}, we show the polarization degree, $\Pi$, as a function of the viewing angle, $\theta_{\rm obs}$. The solid lines are obtained by integrating Eqs.~\eqref{stokes1}-\eqref{stokes3} numerically. We used the jet model of Lyubarsky \cite{Lyubarsky2009}, and evaluated the electromagnetic fields in $R=R_0$, where the edge of the jet is located. The dashed lines correspond to our analytical approximation (i.e.,~Eq.~\ref{polcyl} for $0<q<1/2$, and Eq.~\ref{polparab} for $1/2<q<1$).

In a nearly cylindrical jet where $q=0.3$ (Fig.~\ref{fig:q03}), the analytical approximation works for viewing angles $\theta_{\rm obs} \ll 1/\Gamma_0$, when $\Pi\ll\Pi_{\rm max}$. The approximation overestimates the polarization degree by a factor of a few when $\Pi\sim\Pi_{\rm max}$. In a nearly parabolic jet where $q=0.7$ (Fig.~\ref{fig:q07}), the approximation works for all the viewing angles relevant for blazars, namely $\theta_{\rm obs} \lesssim 1/\Gamma_0$.

\bibliographystyle{apsrev4-2}
\bibliography{main}

\end{document}